\documentclass{article}
\usepackage{spconf,amsmath,graphicx}

\usepackage{bm}
\usepackage{booktabs}
\usepackage{hyperref}
\usepackage{amssymb}
\usepackage{array}
\usepackage{multicol}
\usepackage{multirow}
\usepackage{stfloats}
\usepackage{threeparttable}

\title{Pyramid U-Net for Retinal Vessel Segmentation }

\name{Jiawei Zhang$^{1,3}$, Yanchun Zhang$^{2,3}$, Xiaowei Xu$^4$}
\address{
$^1$ 
School of Computer Science, Fudan University, Shanghai, China \\
$^2$ College of Engineering and Science, Victoria University, Melbourne, Australia \\
$^3$ Cyberspace Institute of Advanced Technology, Guangzhou University, Guangzhou, China \\
$^4$  
Guangdong Cardiovascular Institute, 
Guangdong Provincial People's Hospital, 
Guangzhou, China
}
\begin{document}
%
\maketitle
\begin{abstract}
Retinal blood vessel can assist doctors in diagnosis of eye-related diseases such as diabetes and hypertension, and its segmentation is particularly important for automatic retinal image analysis. 
However, it is challenging to segment these vessels structures, especially the thin capillaries from the color retinal image due to low contrast and ambiguousness. 
In this paper, we propose pyramid U-Net for accurate retinal vessel segmentation.
In pyramid U-Net, the proposed pyramid-scale aggregation blocks (PSABs) are employed in both the encoder and decoder to aggregate features at higher, current and lower levels.
In this way, coarse-to-fine context information is shared and aggregated in each block thus to improve the location of capillaries.
To further improve performance, two optimizations including pyramid inputs enhancement and deep pyramid supervision are applied to PSABs in the encoder and decoder, respectively.
For PSABs in the encoder, scaled input images are added as extra inputs.
While for PSABs in the decoder, scaled intermediate outputs are supervised by the scaled segmentation labels.
Extensive evaluations show that our pyramid U-Net outperforms the current state-of-the-art methods on the public DRIVE and CHASE-DB1 datasets.

\end{abstract}
\section{Introduction}
\label{sec:intro}

Visible structure of retinal vascular may indicate many diseases.
Accurate segmentation helps capture visible changes of retinal vascular structures, which assists doctors in diagnosis of eye related diseases \cite{vessel4,new3,new4}.
Thus, it is particularly important in current retinal image analysis tasks.
For example, 
hypertensive retinopathy is a retinal disease caused by hypertension, and patients with hypertension can find increased vascular curvature or stenosis \cite{vessel4}. 
Conventionally, manual segmentation is performed by experts, which is laborious, time-consuming, and suffers from subjectivity among experts.
Automatic segmentation methods are highly demanded in clinical practice to improve efficiency as well as reliability and reduce the workload of doctors \cite{xu2018quantization}.

In practice, it is difficult to segment vessel structures such as thin capillaries well from the color retinal image, due to low contrast and ambiguousness.
Profiting from the aggregation of multi-scale context information, a variety of deep neural networks \cite{cbi,dcan,new1,new2} have boosted medical image segmentation with a large margin, especially for small objects. 
For example, \cite{mnfn} scaled vessel images to both high and low resolutions, and employed two processes to extract and aggregate feature presentations at multiple scales.  
%
%
\cite{densepooling} integrated feature maps at all or most scales from the previous layers in the encoder sub-network to strengthen feature propagation and encourage feature reuse.
%
%
However, current multi-scale dense connections can not ensure containing both higher-level and lower-level features, which may not fully explore and utilize the information with multiple scales.
For example, the multi-scale dense connections can not supply the higher-level feature maps in the encoder sub-network for accurate localization, because all previous feature maps own a larger size than current layers.
Meanwhile, multi-scale inputs also attract more and more attention in medical image segmentation.
For example, 
MIMO-Net  \cite{mimo} trained the network parameters using multiple resolutions of the input image connecting the intermediate layers for better localization.
MILD-Net \cite{mild} proposed a fully convolutional neural network that counters the loss of information caused by max-pooling by re-introducing the original image at multiple points within the network.
The above methods demonstrate that multi-scale input and deep supervision can efficiently improve the segmentation.

\begin{figure*}[]
	\centering
	\includegraphics[width=0.95\linewidth]{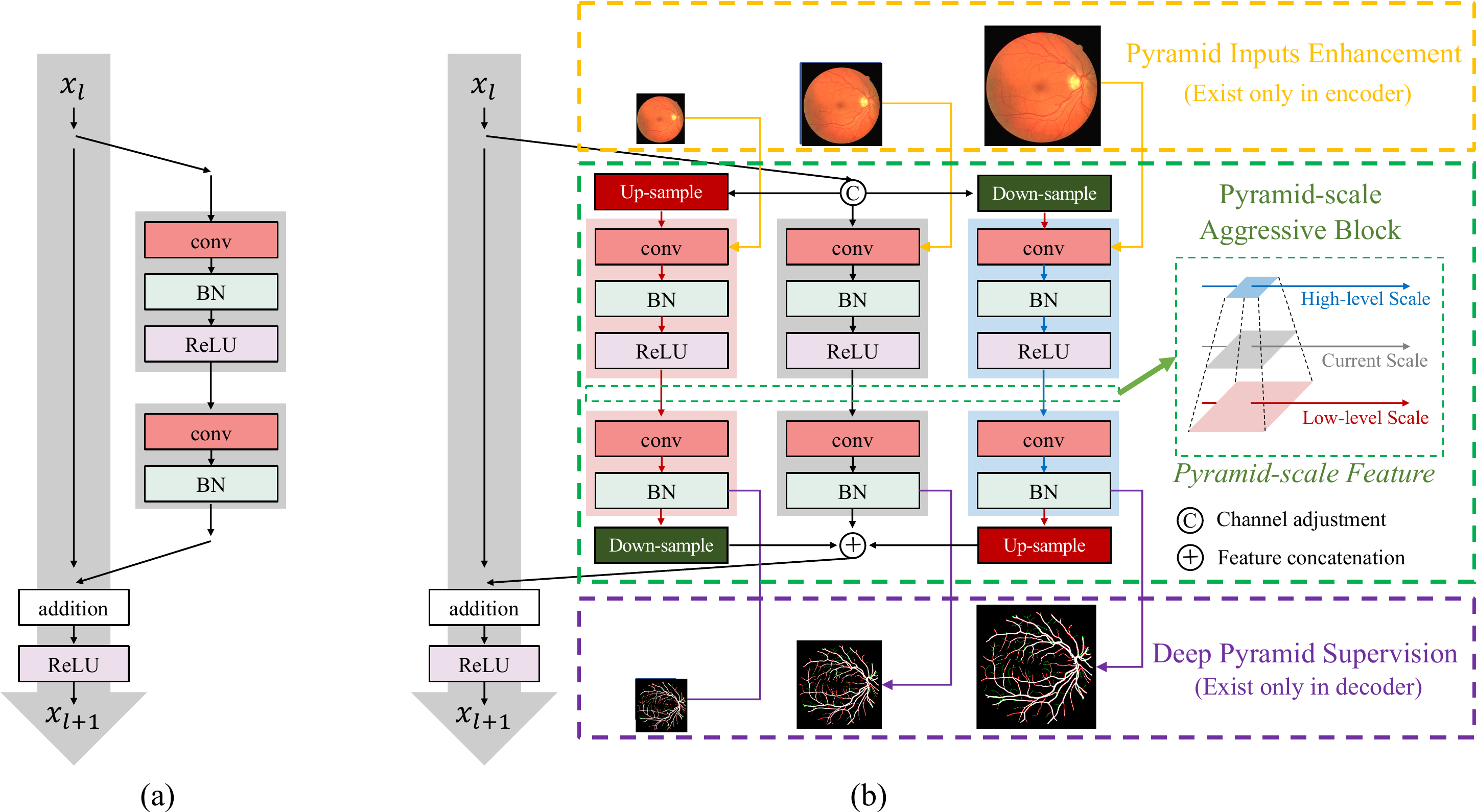}
	\caption{ 
		Illustration of (a) ResNet blocks and (b) our pyramid-scale aggregation blocks (PSABs). 
		PSABs (green rectangle) not only aggregate features with current scale but also from both the higher-level scale and lower-level scales containing coarse-to-fine context information. 
		Meanwhile, pyramid input enhancement (yellow rectangle) and deep pyramid supervision (purple rectangle) are employed to fuse the inputs with corresponding scales and optimize the various-scale features learned in PSABs.%
	}
	\label{fig:msblock}
\end{figure*}

Motivated by the above discovery, in this paper, we propose pyramid U-Net for accurate retinal vessel segmentation.
In pyramid U-Net, the proposed pyramid-scale aggregation blocks (PSABs) are employed in both the encoder and decoder to aggregate features at higher, current and lower levels.
In this way, coarse-to-fine context information is shared and aggregated in each scale thus to improve the location of capillaries. 
To further improve performance, two optimizations including pyramid inputs enhancement and deep pyramid supervision are applied to PSABs in the encoder and decoder, respectively.
For PSABs in the encoder part, scaled input images are added into corresponding blocks as extra inputs to counter the information loss from the pooling layer.
While for PSABs in the decoder part, scaled intermediate outputs are supervised by the scaled segmentation labels  to optimize hierarchical representations and accelerate the training process.
We have conducted comprehensive experiments on two retinal vessel image segmentation datasets including DRIVE \cite{drive} and CHASE-DB1  \cite{chasedb1} with various segmentation networks including U-Net \cite{unet}, DeepVessel \cite{deepvessel}, CE-Net \cite{cenet}. 
The experimental results show that our method can significantly improve the segmentation and achieves state-of-the-art performance on the above two public datasets. 

Our contributions in this paper can be summarized as follows.
\textbf{1)} We introduce the pyramid-scale aggregation blocks (PASBs), which aggregate features at higher, current and lower levels to improve the segmentation performance.
\textbf{2)} 
We propose pyramid input enhancement and deep pyramid supervision to further boost the performance.
The former can reduce the loss of information caused by re-scaling in the encoder, while the latter can deeply supervise the learning process in the decoder.
\textbf{3)} We conducted comprehensive experiments on two public datasets, and experimental results show that our methods can achieve state-of-the-art performance on both datasets.

\section{Methods}

\begin{figure}[]
	\centering
	\includegraphics[width=1\linewidth]{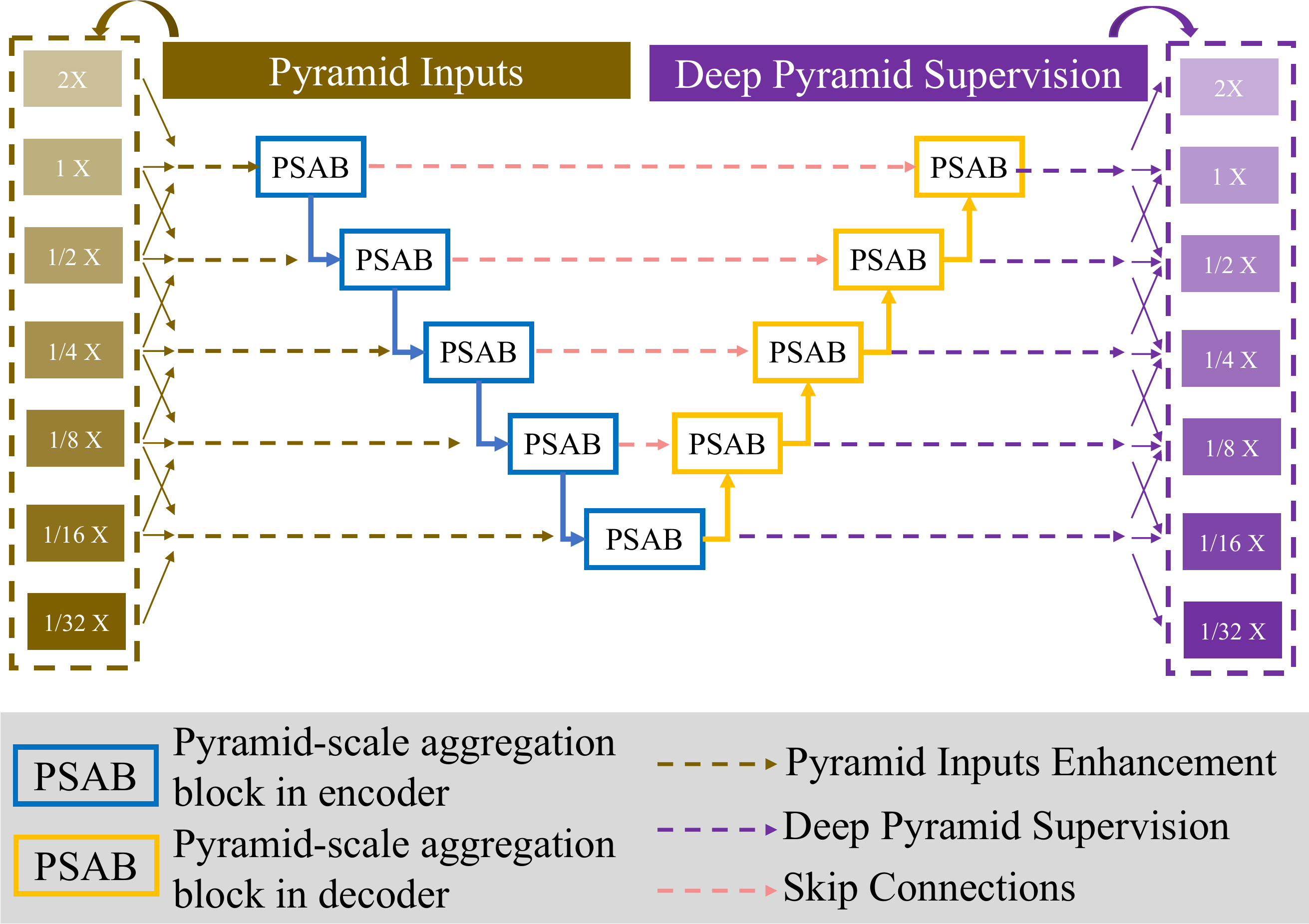}
	\caption{ 
		Structure of the pyramid U-Net.
		%
	}
	\label{fig:overall}
\end{figure}

In this section, the details of pyramid U-Net are presented as shown in Fig. \ref{fig:overall}. We first introduce pyramid-scale aggregation block, and then describe two optimizations including pyramid input enhancement and deep pyramid supervision.

\subsection{Pyramid-scale Aggregation Block}
Pyramid-scale aggregation blocks (PSABs) are based on the widely adopted ResNet block \cite{res}.
The structure of ResNet block \cite{res} is illustrated in the dashed box in Fig. \ref{fig:msblock}, which is defined as
$$ X_{l+1}=f(X_{l}) + X_{l},  \eqno{(1)}$$
where $X_{l}$ and $X_{l+1}$ are the input and output of the current layer, while $f(\cdot)$ represents the learning process of the current layer. 
Fig. \ref{fig:msblock} illustrates the detailed structure of traditional ResNet blocks and our PSABs.
Different from ResNet blocks, PSABs perform processing at three parallel pyramid scales including the higher, current and lower scales.
In each scale, the processing steps are almost the same as that in traditional ResNet blocks.
Some extra steps such as up-sampling and down-sampling are adopted at higher and lower scales to adjust scales.
%
In order to reduce the potential increase of computational cost, the number of channels of the input $X_{l}$ has been reduced to half, while the number of channels of resized inputs $X_{l}^p$ and $X_{l}^d$ are reduced to one-fourth.
The outputs of channel adjustment are fed to the processing steps at three scales and are processed in parallel.
The three outputs at higher, current and lower scales are then concatenated. 
The whole process is formulated as follows,

$$ \widetilde X_{l+1}= H( f(\widehat X_{l}^{p}), f(\widehat X_{l}), f(\widehat X_{l}^{d}))  + X_{l},  \eqno{(2)}$$
where $X_{l}^{p}$ and $X_{l}^{d}$ are the up-sampled and down-sampled results of the current input $X_{l}$ with channel adjustment, respectively.
Meanwhile, $\widehat X_{l}^{p}$, $\widehat X_{l}$ and $\widehat  X_{l}^{d}$ are the enhancement results by the pyramid input enhancement, which is detailed in section 2.2. 
$H(\cdot)$ represents the aggregation process, which performs re-scaling and feature concatenation.
%
$\widetilde X_{l+1}$ is the strengthened results of $ X_{l+1}$ by PASB.

To improve the efficiency of feature extraction, we also employ an attention mechanism \cite{senet,li2} in PSAB as follows,
$$
\bm{\Phi} (  \mathbf{ \widetilde X }_{l+1} ) =
\mathbf{W} ( \Phi_{Avg} (\mathbf{\widetilde X}_{l+1}  ) ) +
\mathbf{W} ( \Phi_{Max }(\mathbf{\widetilde X }_{l+1}  ) ).
\quad \eqno{\left(3\right)}
$$
$$
\bm{\Psi} (  \mathbf{ \widetilde X }_{l+1} ) = 
\bm{\sigma} (
\bm{\Phi} (  \mathbf{ \widetilde X }_{l+1} ) 
\otimes \mathbf{\widetilde X}_{l+1}.
\quad \eqno{\left(4\right)}
$$
where $\bm{\Psi}(\cdot)$ is the operation of attention process, $\mathbf{W}$ is the conventional operation using 1$\times$1 kernels for channel adjustment, and $\bm{\sigma}$ is the activation function. 
Average-pooling $\Phi_{Avg}(\cdot)$ and max-pooling $\Phi_{Max}(\cdot)$ are adopted for aggregating channel information.

\subsection{Pyramid Input Enhancement}

Pyramid input enhancement introduces the input image to PSABs to enlarge feature fusion.
Particularly, the input image is scaled at higher, current and lower scales which are fed to three parallel processing steps at different scales in the PSAB to reduce the loss of information caused by scaling.
The above three pyramid-scale images are concatenated with corresponding outputs of up-sampling, down-sampling and channel adjustment, respectively.
For clearly, $X_{l}$ is denoted as the input of the current layer, and $X_{l}^{p}$ and $X_{l}^{d}$ are results at higher and lower scales, respectively.
Meanwhile, $I_{l-1}$, $I_{l}$ and $I_{l+1}$ are the corresponding scaled inputs of $X_{l}^{d}$, $X_{l}$ and $X_{l}^{p}$, which have the same size. 
The fusion process of the current scale is formulated as follows,

$$
\widehat X_{l-1}=H(X_{l}^{d} , \mathbf{W}^{d} (I_{l-1}) ),
\eqno{(4)}
$$
$$
\widehat X_{l}=H(X_{l} , \mathbf{W} (I_{l}) ),
\eqno{(5)}
$$
$$
\widehat X_{l+1}=H(X_{l}^{p} , \mathbf{W}^{p} (I_{l+1}) ),
\eqno{(6)}
$$
where $\mathbf{W}^{p}(\cdot),\mathbf{W}^{d}(\cdot)$ and $\mathbf{W}(\cdot)$ represents 3$\times$3 convolutional operations and is applied before concatenating to the features with pyramid-scale, and $H(\cdot)$ denotes the channel adjustment.

\subsection{Deep Pyramid Supervision}

In order to optimize hierarchical representations from multiple scales, deep pyramid supervision is further adopted. 
To realize deep pyramid supervision, the learned feature maps at multiple scales from each PSAB in the decoder is fed into a plain 3 $\times$ 3 convolutional layer followed by a sigmoid function.
The deep pyramid supervision at the $l$th scale of the decoder can be defined as,
$$
L_{l} = L(Y_{l}^{p},M_{l-1}) + L( Y_{l},M_{l}) + L( Y_{l}^{d},M_{l+1}).
\eqno{(7)}
$$
The ground truths $M$ are scaled to the same size of the pyramid-scale feature maps for deep supervision, e.g., $ Y_{l}^{p}, Y_{l}$ and $ Y_{l}^{d}$ are supervised by the corresponding ground truth $M_{l-1}$, $M_{l}$ and $M_{l+1}$, respectively.
Each of the auxiliary feature masks is followed by a dropout layer (set to 50\%) and a convolutional layer followed by a Softmax layer
to get the auxiliary outputs.
Our loss combines cross-entropy loss and Intersection over Union (IoU) loss.

\section{Experiments}

\subsection{Datasets}

We used two public available retinal vessel datasets, DIRVE \cite{drive} and CHASE-DB1 \cite{chasedb1} for evaluation.
The DRIVE dataset contains 40 images with a resolution of 565 $\times$ 584 pixels, which were acquired using a Canon CR5 non-mydriatic 3CCD camera with a 45-degree field of view (FOV).
Meanwhile, the FOV of each image is circular with a diameter of approximately 540 pixels, the images have been cropped around the FOV.
The images are resized to 448 $\times$ 448 pixels.
The set of 40 images has been divided into a training and a test set, both containing 20 images. 
In particular, we pre-train our network on PASCAL VOC 2012 to       alleviate potential over-fitting due to the limited dataset for retinal vessel segmentation. 
The CHASE-DB1 dataset consists of 28 vascular patch images with a resolution of 999 $\times$ 960.  
Following the configuration in \cite{chasedbd}, we use the first 20 images as the training set and the remaining 8 images as the test set.

\subsection{Implementations and Evaluation}
We implemented all experiments in PyTorch platform and trained models on an Nvidia GeForce Titan X machine
with 12 GB memory.
Meanwhile, we employed CE-Net as our backbones to implement PSABs, pyramid input enhancement and deep pyramid supervision.
During training, we adopted Adaptive Moment Estimation (Adam) with a batch size of 4 and a weight decay of 0.0001.
Training techniques like data argumentation (horizontal flip, vertical flip and diagonal flip) and learning rate decay are equipped.
To evaluate our model, we used Sensitivity (Sen), Specificity (Spec), Accuracy (Acc) and Area Under the ROC Curve (AUC) as evaluation metrics. 
The aforementioned metrics are calculated as follows:
$
\operatorname{Sen}=\text{TP}/(\text{TP}+\text{FN}),\quad{\text{Spec}}=\text{TN}/(\text{TN}+\text{FP}),\quad{\text{Acc}}=(\text{TP}+\text{TN})/(\text{TP}+\text{TN}+\text{FP}+\text{FN}).
$
If pixels are correctly classified to objects or backgrounds, they will be annotated as True Positive (TP) or True Negatives (TN), respectively. 
Meanwhile, pixels are misclassified to objects or backgrounds are labeled as False Positive (FP) or False Negatives (FN), respectively.

\begin{table}[h]
	\centering
	\caption{ Performance comparison of pyramid U-Net with state-of-the-art methods on the DRIVE dataset. 
	}
	\vspace{0.2cm}
	\label{tab:DRIVE}
	\begin{tabular}{p{3.5cm}<{\centering}||p{0.7cm}<{\centering}p{0.7cm}<{\centering}p{0.7cm}<{\centering}p{0.7cm}<{\centering}}
		\toprule[2pt]
		\multirow{2}{*}{Method} & 
		 \multirow{2}{*}{Sens}  & \multirow{2}{*}{Spec}&\multirow{2}{*}{Acc}& \multirow{2}{*}{AUC}   \\ 
		&     &     &                    \\    \hline \hline

		\small{Azzopardi \cite{azzopardi} (2015)}      \quad       &0.7655 &0.9704  &0.9442       &0.9614  \\
        \small{Roychowdhury \cite{Roychowdhury} (2015) }        &0.7395 &0.9782    &0.9494   &0.9672  \\
        \small{U-Net    \cite{unet} (2015)   }     &0.7531  &0.9645  &0.9445       &0.9601      \\
        \small{DeepVessel\cite{deepvessel} (2016) }  &0.7612 &0.9768        &0.9523        &0.9752     \\
        \small{CE-Net  \cite{cenet} (2019) }      &\textbf{0.8309}   &0.9747  &0.9545        &0.9779   \\
        \small{\textbf{Pyramid U-Net}}  &0.8213   &\textbf{0.9807}  &\textbf{0.9615}        &\textbf{0.9815}   \\

        \bottomrule[2pt]
		
	\end{tabular}
\end{table}


\subsection{\textbf{Comparison with State-of-the-art Works}} 
\begin{figure}[h]
	\centering
	\includegraphics[width=0.99\linewidth]{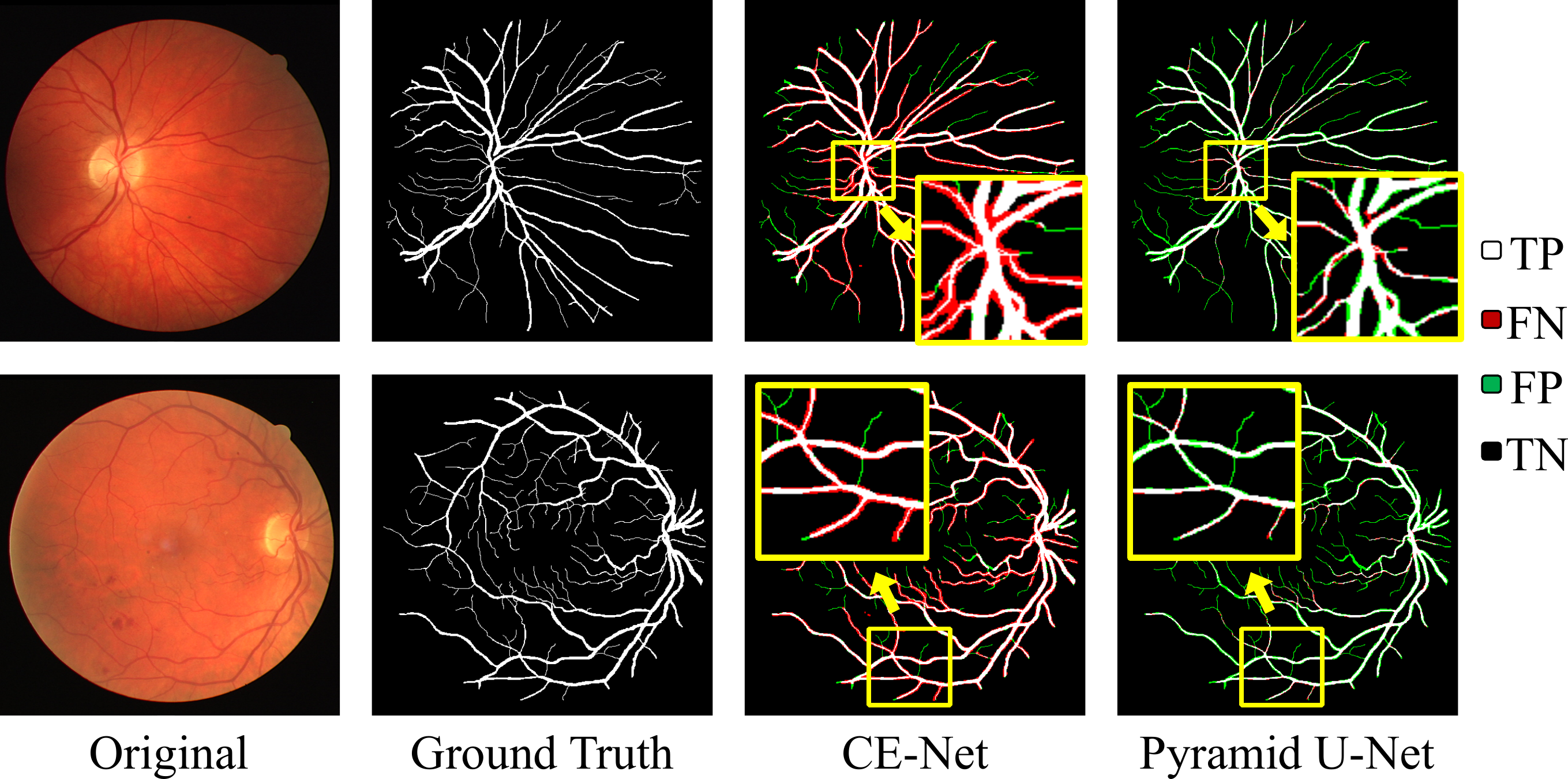}
	\caption{ 
		Visual comparison of CE-Net \cite{cenet} and our pyramid U-Net
		with corresponding original images and ground truths
		on the DRIVE dataset. 
		%
	}
	\label{fig:result}
\end{figure}

We compared our pyramid U-Net with existing state-of-the-art works including U-Net \cite{unet}, DeepVessel \cite{deepvessel} and  CE-Net \cite{cenet} on DRIVE and DECHASE1 datasets.
The results are summarized in Table \ref{tab:DRIVE} and Table \ref{tab:CHASEDB}.
On the DRIVE dataset, compared with the state-of-the-art method CE-Net, pyramid U-Net obtains a higher performance on Spec (improved by 0.6\%), Acc (improved by 0.7\%), AUC (improved by 0.36\%) and a competitive result on Sen, which is only 0.32\% worse.
Meanwhile, our proposed pyramid U-Net outperforms state-of-the-art methods on all metrics on the CHASE-DB1 dataset.
The visual comparison is shown in Fig. \ref{fig:result}. 
White and black pixels are correct predictions of object and background, respectively, while red and green pixels are incorrect predictions.  
We can notice that our proposed pyramid U-Net evidently improves the segmentation performance.
Especially for those narrow, low-contrast and ambiguous retinal vessels, which are highlighted by yellow rectangles.

\begin{table}[tp]
	\centering
	\caption{ Performance comparison of pyramid U-Net with state-of-the-art methods on the CHASE-DB1 dataset. 
	}
	\vspace{0.2cm}
	\label{tab:CHASEDB}
	\begin{tabular}{p{3.5cm}<{\centering}||p{0.7cm}<{\centering}p{0.7cm}<{\centering}p{0.7cm}<{\centering}p{0.7cm}<{\centering}}
		\toprule[2pt]
		\multirow{2}{*}{Method} & 
		 \multirow{2}{*}{Sens}  & \multirow{2}{*}{Spec}&\multirow{2}{*}{Acc}& \multirow{2}{*}{AUC}   \\ 
		&     &     &                    \\    \hline \hline
		
        \small{Azzopardi \cite{azzopardi} (2015) }     &0.7585 &0.9587  &0.9387       &0.9487  \\

        \small{U-Net \cite{unet} (2015) }	& 	0.7675& 	0.9631& 	0.9409&	0.9705\\
        \small{DeepVessel \cite{deepvessel} (2016) }& 0.7412& 	0.9701&	0.9609&	0.9790\\
        \small{CE-Net \cite{cenet} (2019) }	& 0.7841& 	0.9725& 	0.9583& 	0.9787\\
        \small{\textbf{Pyramid U-Net} }	& \textbf{0.8035}& 	\textbf{0.9787}& 	\textbf{0.9639}& 	\textbf{0.9832}\\
        \bottomrule[2pt]
		
	\end{tabular}
\end{table}

\section{Conclusion}

In this paper, we proposed pyramid U-Net for retinal vessel segmentation.
In pyramid U-Net, the proposed pyramid-scale aggregation blocks (PSABs) are employed in both the encoder and decoder to aggregate features at higher, current and lower levels.
To further improve performance, two optimizations including pyramid inputs enhancement and deep pyramid supervision are applied to PSABs in the encoder and decoder, respectively.
We have conducted comprehensive experiments on two retinal vessel image segmentation datasets including DRIVE \cite{drive} and CHASE-DB1  \cite{chasedb1}.
Experimental results show that our pyramid-scale aggregation block can efficiently improve the segmentation performance.

\section{Acknowledgments}
This work is supported by the National Natural Science Foundation of China (No.62006050).

\vfill\pagebreak

\bibliographystyle{IEEEbib}
\bibliography{egbib}

\end{document}